\documentclass[12pt,a4paper,twoside]{article}

\usepackage{graphicx}
\usepackage{epic}
\usepackage{epstopdf}
\usepackage{amssymb}
\usepackage{amsmath}
\usepackage{sectsty}
\usepackage{xcolor,pict2e}

\usepackage[utf8x]{inputenc} 
\usepackage[english]{babel}

\def\abs#1{\left\lvert#1\right\rvert}

\sectionfont{\huge}

\begin{document}

\begin{center}
{\huge\bf Cybernetic Interpretation of the Riemann Zeta Function}
\end{center}

\vspace{0.1cm}
\begin{center}
{\it Petr Kl\'{a}n, Dept. of System Analysis, University of Economics in Prague, Czech Republic, petr.klan@vse.cz}
\end{center}

\vspace{0.3cm}
\noindent
{\bf Abstract:} This paper uses cybernetic approach to study behavior of the Riemann zeta function. It is based on the elementary cybernetic concepts like feedback, transfer functions, time delays, PI (Proportional--Integral) controllers or FOPDT (First Order Plus Dead Time) models, respectively. An unusual dynamic interpretation of the Riemann zeta function is obtained.

\vspace{0.7cm}

\section{Introduction}

\vspace{0.3cm}\noindent {\it Zeta} function \cite{edw} is associated with equation

\begin{equation}
 \sum_{n=1}^{\infty}\dfrac{1}{n^s}=\prod_{p=2,3,5 \ldots}\dfrac{1}{1-p^{-s}} \label{zeta}
\end{equation}
introducing the fundamental relationship between natural numbers $n = 1,2,3,4,5 \ldots$ and prime numbers $p = 2,3,5,7,11 \ldots $ for real $s>1$. It was discovered by L. Euler in 1737. A full idea of this relationship excels at expansion to the individual base elements

$$
\dfrac{1}{1^s}+\dfrac{1}{2^s}+\dfrac{1}{3^s}+\dfrac{1}{4^s}+\dfrac{1}{5^s}+\ldots =\dfrac{1}{1-2^{-s}}\,\dfrac{1}{1-3^{-s}}\,\dfrac{1}{1-5^{-s}}\,\dfrac{1}{1-7^{-s}}\,\dfrac{1}{1-11^{-s}}\ldots,
$$    
in which an infinite summation over all the natural numbers (left side) and an infinite product over all the prime numbers (right side) are contained. L. Euler assumed the parameter $s$ on both sides as a real number. Equation (\ref{zeta}) is considered to be an input to the mystery of interrelation between natural and prime numbers that is waiting for hundreds of years to be discovered \cite{cla}. \\ 
\\
The infinite sum $\sum 1/n^s$ from (\ref{zeta}), typically for $s>1$, is called {\it zeta function} $\zeta(s)$, thus

\begin{equation}
\zeta(s)= \sum_{n=1}^{\infty}\dfrac{1}{n^s}. \label{zetale}
\end{equation}
Function diverges for $s<1$ and converges for $s>1$. For $s=1$, the function $\zeta(s)$ is projected to a harmonic series, which is divergent as well \cite{kla}.\\
\\
G. F. Riemann proposes a specific analytic function in the field of all complex numbers $s=a+\rm{i}b$ in his work {\it On the Number of Prime Numbers Under a Given Magnitude} (\cite{edw}, Riemann's original work is a part of this publication)  that coincides with zeta function for Re$(s)> 1$. This new function is called the {\it Riemann zeta function} and denoted in the same way as zeta function (\ref{zetale}). Whilst coincident with (\ref{zetale}) introduced by L. Euler for values of $a>1$ and $b=0$, it is more formally considered for values of $a<1$. The Riemann zeta function stays in the effect throughout the interval of $s$, including $a<1$. In considering language of the complex analysis, the Riemann zeta function represents a meromorphic function in whole complex plane except point $s = 1$, where it has no final value \cite{kri}\footnote {This article provides an elegant proof of equation (\ref{zeta}) for Re$(s)>1$.}. It greatly contributes to the analysis of prime numbers distribution \cite{cla,edw,ste}.\\
\\
Riemann zeta function $\zeta(s)$ has two sorts of zero values\footnote {Terms like zeros and poles of  transfer functions are well known in cybernetics too \cite {klagor}.} or values $s$ for which $\zeta (s)=0$. Some of them, for example $s=0.5 + \rm{i}14.135$,  $s=0.5+ \rm{i}21.022$, $s=0.5+ \rm{i} 25.011$, $s =0.5+\rm{i} 30.425$, $s = 0.5 + \rm{i} 32.935 \ldots$ are called {\it nontrivial}. One can ask, where do all the complex zero values of the Riemann zeta function locate. As extremely interesting, it is taken the area of complex plane determined by interval $0 \leq a \leq 1$ for any real $b$ which is called critical strip, see Fig. \ref{cplane}. The Riemann zeta function has zero values in negative integers $\zeta(-2) = 0,$ $\zeta(-4) = 0,$ $\zeta(-6) = 0$ \ldots as well. They are called {\it trivial}. The Riemann hypothesis conjectures that all the critical strip nontrivial zeros lie on midst critical line of $1/2$. In other words, all the zero values of the Riemann zeta function (except trivial) are assumed to receive values of $s = 0.5 + \rm{i}b$ or $1/2+\rm{i}b$ for real $b$. When finding a single zero value in the critical strip outside this midst line, Riemann hypothesis fails. The hypothesis has not been proven yet. Due to its fundamental significance, it belongs to the millennium problems \cite{cla}.

{\small
\begin{figure}[htbp]\begin{center}
\setlength{\unitlength}{1.0mm}
\begin{picture}(130,115)
\thicklines

\put(5,60){\line(1,0){100}}
\put(20,10){\line(0,1){100}}
\put(80,10){\line(0,1){100}}

\thinlines

\put(50,10){\line(0,1){100}}

\put(50, 74.1){\circle*{1.5}}
\put(57,74.1){\makebox(0,0){\footnotesize $14.1$}}
\put(50, 45.9){\circle*{1.5}}
\put(58,45.9){\makebox(0,0){\footnotesize $-14.1$}}

\put(50, 81){\circle*{1.5}}
\put(57,81){\makebox(0,0){\footnotesize $21.0$}}
\put(50, 39){\circle*{1.5}}
\put(58,39){\makebox(0,0){\footnotesize $-21.0$}}

\put(50, 85){\circle*{1.5}}
\put(57,85){\makebox(0,0){\footnotesize $25.0$}}
\put(50, 35){\circle*{1.5}}
\put(58,35){\makebox(0,0){\footnotesize $-25.0$}}

\put(50, 90.4){\circle*{1.5}}
\put(57,90.3){\makebox(0,0){\footnotesize $30.4$}}
\put(50, 29.6){\circle*{1.5}}
\put(58,29.7){\makebox(0,0){\footnotesize $-30.4$}}

\put(50, 92.9){\circle*{1.5}}
\put(57,93.3){\makebox(0,0){\footnotesize $32.9$}}
\put(50, 27.1){\circle*{1.5}}
\put(58,26.7){\makebox(0,0){\footnotesize $-32.9$}}

\put(50, 97.6){\circle*{1.5}}
\put(57,97.6){\makebox(0,0){\footnotesize $37.6$}}
\put(50, 22.4){\circle*{1.5}}
\put(58,22.4){\makebox(0,0){\footnotesize $-37.6$}}

\put(50, 100.9){\circle*{1.5}}
\put(57,100.9){\makebox(0,0){\footnotesize $40.9$}}
\put(50, 19.1){\circle*{1.5}}
\put(58,19.1){\makebox(0,0){\footnotesize $-40.9$}}

\put(18,57){\makebox(0,0){$0$}}
\put(46,57){\makebox(0,0){$1/2$}}
\put(78,57){\makebox(0,0){$1$}}

\put(17,70){\makebox(0,0){$10$}}
\put(17,80){\makebox(0,0){$20$}}
\put(17,90){\makebox(0,0){$30$}}
\put(17,100){\makebox(0,0){$40$}}

\put(15.5,50){\makebox(0,0){$-10$}}
\put(15.5,40){\makebox(0,0){$-20$}}
\put(15.5,30){\makebox(0,0){$-30$}}
\put(15.5,20){\makebox(0,0){$-40$}}

\put(20,3){\line(0,1){5}}
\put(80,3){\line(0,1){5}}
\put(50,4){\vector(1,0){30}}
\put(50,4){\vector(-1,0){30}}
\put(49,6){\makebox(0,0){\footnotesize Critical strip}}

\put(67,68){\makebox(0,0){\footnotesize Critical line}}
\put(55,68){\vector(-1,0){5}}
\end{picture}
\caption{\small Critical strip of complex plane showing positions of nontrivial zero values.} \label{cplane}
\end{center}
\end{figure}
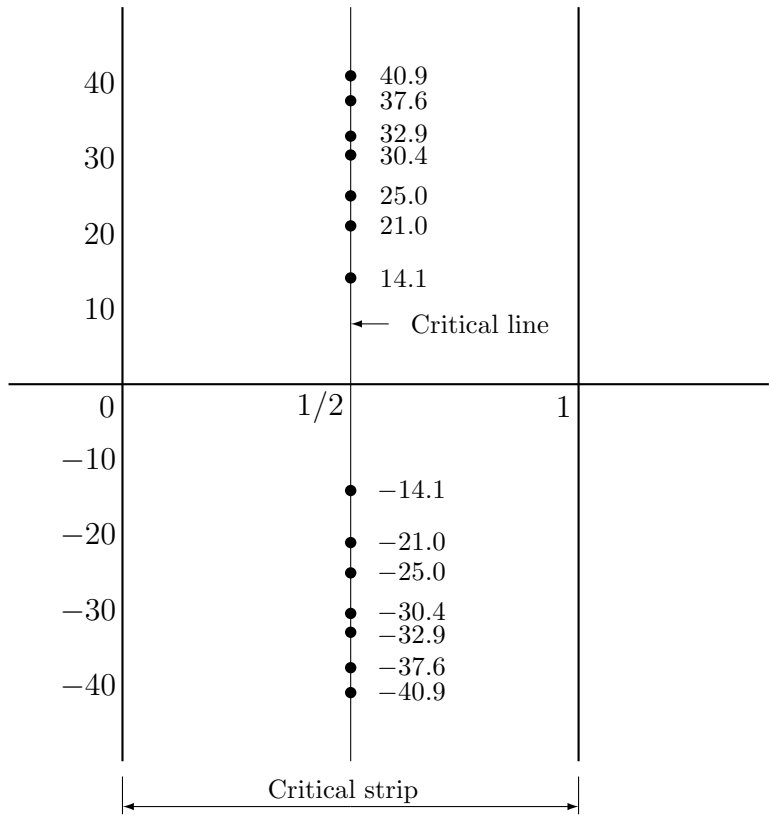}

\vspace{-0.7cm}\noindent The paper shows that for $a > 1$, the Riemann zeta function can be interpreted by an endless serial connection of the elementary cybernetic concepts like PI controllers or FOPDT models \cite{klagor}. It is shown that this interpretation brings a new perspective to the Riemann zeta function with possible impacts to the Riemann hypothesis. 

\section{Single transfer functions}

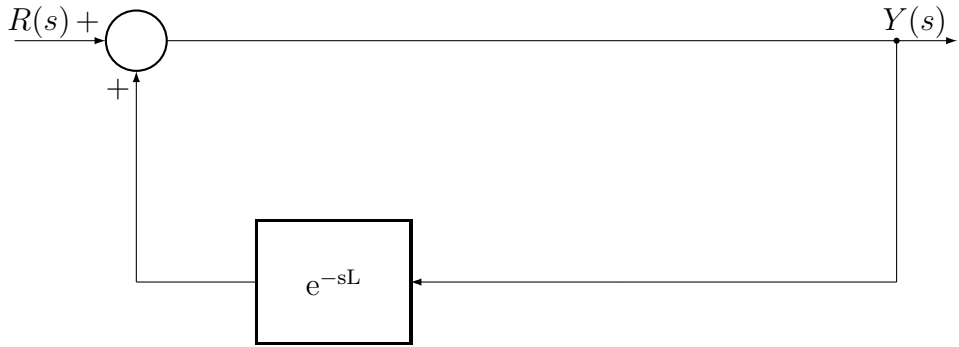
\begin{figure}[htb] 
\begin{center}
\setlength{\unitlength}{0.8mm}
\begin{picture}(140,70)
\thicklines 
\put(30,10){\framebox(25,20){$\rm{e}^{-sL}$}}

\put(10, 60){\circle{10}}

\thinlines

\put(10,20){\line(1,0){20}} \put(90,20){\vector(-1,0){35}}
\put(135,20){\line(-1,0){45}}\put(-10,60){\vector(1,0){15}}

\put(10,20){\vector(0,1){35}} \put(55,60){\line(1,0){60}}
\put(115,60){\vector(1,0){30}}\put(15,60){\line(1,0){40}}
\put(135,20){\line(0,1){40}}\put(135, 60){\circle*{1}}

\put(-6,63){\makebox(0,0){$R(s)$}}\put(138,63){\makebox(0,0){$Y(s)$}}
\put(2,63){\makebox(0,0){$+$}}\put(7,52){\makebox(0,0){$+$}}
\end{picture}
\caption{Base element (\ref{elemente}) represented by the transfer function.} \label{elementL}
\end{center}
\end{figure}

\vspace{0.3cm}\noindent The right--hand side of equation (\ref{zeta}) is composed of an infinite product of base elements 

\begin{equation}
\dfrac{1}{1-p^{-s}}, \label{elementp}
\end{equation}
where $p$ is a prime and $s$ a complex number. Cybernetic considerations frequently use the base of natural logarithm e unlike primes $p$ in the number theory. Thus, rewrite the base element (\ref{elementp}) equivalently to

\begin{equation}
\dfrac{1}{1-\rm{e}^{-s \ln p}}. \label{elemente}
\end{equation}
If $\rm{e}^x=p^{-s}$ for a certain $x$, then $x=-s\ln p$. Now, the infinite product (\ref{zetale}) becomes

\begin{equation}
\prod_{p}\dfrac{1}{1-\rm{e}^{-s\ln p}}=\dfrac{1}{1-\rm{e}^{-s\ln 2}}\,\dfrac{1}{1-\rm{e}^{-s\ln 3}}\,\dfrac{1}{1-\rm{e}^{-s\ln 5}}\,\dfrac{1}{1-\rm{e}^{-s\ln 7}}\,\dfrac{1}{1-\rm{e}^{-s\ln 11}}\ldots     \label{zetaesp}
\end{equation}
In Laplace transformation \cite{doe}, the expression $\rm{e}^{-s \ln p}$ represents a transfer function including time delay $\ln p$. Generally, $G(s) = \rm{e}^{-sL}$, where $L = \ln p$ \cite {klagor}. Looking through the eyes of transfer functions, a block scheme including positive feedback for the entire expression (\ref{elemente})  is in Fig. \ref{elementL}. Here, $R(s)$ is Laplace transformation of a reference input and $Y(s)$ is Laplace transformation of the corresponding output. In that case $Y(s)/R(s)=1/(1-\rm{e}^{-s \ln p})$.

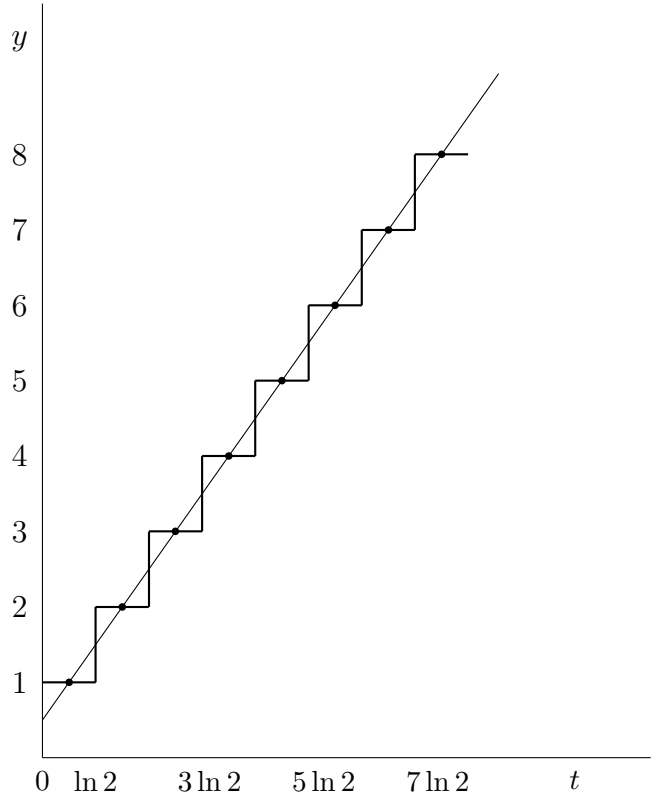
\begin{figure}[htb] 
\begin{center}
\setlength{\unitlength}{1.0mm}
\begin{picture}(100,115)
\thinlines 
\put(10,10){\line(1,0){80}}
\put(10,10){\line(0,1){100}}

\put(10,15){\line(7,10){60}}

\thicklines
\put(10,20){\line(1,0){7}}
\put(17,20){\line(0,1){10}}
\put(17,30){\line(1,0){7}}
\put(24,30){\line(0,1){10}}
\put(24,40){\line(1,0){7}}
\put(31,40){\line(0,1){10}}
\put(31,50){\line(1,0){7}}
\put(38,50){\line(0,1){10}}
\put(38,60){\line(1,0){7}}
\put(45,60){\line(0,1){10}}
\put(45,70){\line(1,0){7}}
\put(52,70){\line(0,1){10}}
\put(52,80){\line(1,0){7}}
\put(59,80){\line(0,1){10}}
\put(59,90){\line(1,0){7}}

\put(13.5, 20){\circle*{1}}
\put(20.5, 30){\circle*{1}}
\put(27.5, 40){\circle*{1}}
\put(34.5, 50){\circle*{1}}
\put(41.5, 60){\circle*{1}}
\put(48.5, 70){\circle*{1}}
\put(55.5, 80){\circle*{1}}
\put(62.5, 90){\circle*{1}}

\thinlines

\put(7,20){\makebox(0,0){$1$}}
\put(7,30){\makebox(0,0){$2$}}
\put(7,40){\makebox(0,0){$3$}}
\put(7,50){\makebox(0,0){$4$}}
\put(7,60){\makebox(0,0){$5$}}
\put(7,70){\makebox(0,0){$6$}}
\put(7,80){\makebox(0,0){$7$}}
\put(7,90){\makebox(0,0){$8$}}

\put(17,7){\makebox(0,0){\small $\ln 2$}}
\put(32,7){\makebox(0,0){\small $3\ln 2$}}
\put(47,7){\makebox(0,0){\small $5\ln 2$}}
\put(62,7){\makebox(0,0){\small $7\ln 2$}}

\put(7,105){\makebox(0,0){$y$}}
\put(80,7){\makebox(0,0){\small $t$}}
\put(10,7){\makebox(0,0){\small $0$}}

\end{picture}
\caption{Step response associated with transfer function $1/(1-\rm{e}^{-s \ln 2})$.} \label{zetat}
\end{center}
\end{figure}

\vspace{-0.3cm}\noindent Reference inputs are usually realized by a unit step $1(t)$. The latter is defined as $0$ for times $t<0$ and $1$ for $ t>0$, in Laplace transformation $1/s$. Assume that $r(t)=1(t)$ is a unit step and determine output response $y(t)$ of transfer function (\ref{elemente}) to a unit step where the signals $r(t)$ and $y (t)$ are denoted by $R(s)$ and $Y(s)$ in the Laplace transformation.\\
\\
Suppose time delay $L=\ln 2$ related to the first prime number in the positive feedback loop of Fig. \ref {elementL}. Corresponding transfer function is $Y(s)/R(s)=1/(1- \rm{e}^{- s \ln 2})$. A reference input unit step is immediately transferred to the output, or $y (t)=1$ for $t>0$. This output is delayed in the feedback and after time delay $L$ it is added to the reference input, or $y (t)=2$ for $t>L$. It again is delayed by time $L$, added to reference input. It results in $y(t) = 3$ for $t> 2L $ etc. A whole response to unit step reference input (transient response) is in Fig. \ref{zetat}. Due to the positive feedback, a regular multistep signal $y(t)$ is obtained\footnote{It agrees with \cite{doe} which introduces the same Laplace transformation for time functions shown in Fig. \ref{zetat}.}.

\vspace{-0.3cm}\noindent Transient response in Fig. \ref{zetat} can not be simply described. Therefore,  an approximation is used in the sequel. The points placed just in the middle of time delays in Fig. \ref{zetat} join a line beginning at $y(0) = 0.5$. Some of these points and the corresponding values of transient response are collected in Tab. \ref{zetaval}. 

\begin{table}[htb]
\begin{center}
\begin{tabular}{c|c|c|c|c|c}
$t$ & $\ln 2/2$ & $3\ln 2/2$ & $5\ln 2/2$ & $7\ln 2/2$ & $\ldots$\\ \hline
$y$ & $1$ & $2$ & $3$ & $4$ & $\ldots$\\
\end{tabular}
\caption{\small A collection of transient pairs associated with transfer function $1/(1-\rm{e}^{-s \ln 2})$.} \label{zetaval}
\end{center}
\end{table} 

\vspace{-0.3cm}\noindent Considering pairs in Tab. \ref{zetaval} one can make the following interesting deduction: the approximated line transient response is described by  

$$
y(t)=0.5+\dfrac{1}{\ln 2} t
$$
which coincides with the transient response of a PI controller given by transfer function

\begin{equation}
\dfrac{1}{2}+\dfrac{1}{sL} \label{pi}
\end{equation}
or $0.5+1/s \ln 2$, where $L=\ln 2$ represents an integration time constant. Notice that the line in Fig. \ref{zetat} corresponds to (\ref{pi}) as well if the first--order Pade approximation is used in the form $\rm{e}^{- s \ln p} = (1-s \ln p/2)/(1 + s \ln p/2)$. After substituting the latter into $1/(1- \rm{e}^{- s \ln p})$ it implies that $1/[1-(1-s\ln p/2)/(1+s\ln p/2)]=1/2+1/(s\ln p)$.\\
\\
Since an infinite product in the Riemann zeta function, it is concluded that the latter can be approximated by an endless series combining transfer functions of PI controllers (\ref{pi})

\begin{equation}
\prod_{p}\left(\dfrac{1}{2}+\dfrac{1}{s \ln p}\right) \label{zetapi}
\end{equation}
or

$$
\left(0.5+\dfrac{1}{s\ln 2}\right)\left(0.5+\dfrac{1}{s\ln 3}\right)\left(0.5+\dfrac{1}{s\ln 5}\right)\left(0.5+\dfrac{1}{s\ln 7}\right)\left(0.5+\dfrac{1}{s\ln 11}\right)\ldots
$$
This approximation is called {\it PI approximation} or simple approximation of the Riemann zeta function. Connection of something fundamental as the Riemann zeta function is in the theory of numbers and something as PI controllers fundamental in cybernetics is really surprising based on a big context difference of both these subjects.

\section{Case $\zeta(\rm{i}b)$}

\vspace{0.3cm}\noindent By virtue of formulation (\ref{zetaesp}) in which $s = \rm{i} b$ for non--negative real $b$ and $a=0$, we can form $\zeta(\rm{i} b)$ by 

$$
\prod_{p}\dfrac{1}{1-\rm{e}^{-\rm{i}b\ln p}}.
$$
Based on (\ref{zetapi}), it results in a PI approximation

\begin{equation}
\zeta_{\rm{PI}}(\rm{i}b)=\prod_{p}\left(\dfrac{1}{2}+\dfrac{1}{\rm{i}b \ln p}\right) \label{zetapif}
\end{equation}
composed of the frequency transfer functions.

\section{Case $\zeta(a+\rm{i}b)$}

\begin{figure}[htb] 
\begin{center}
\setlength{\unitlength}{0.8mm}
\begin{picture}(140,70)
\thicklines 
\put(30,10){\framebox(25,20){$\rm{e}^{-sL}$}}
\put(90,10){\framebox(25,20){$K$}}

\put(10, 60){\circle{10}}

\thinlines

\put(10,20){\line(1,0){20}} \put(90,20){\vector(-1,0){35}}
\put(135,20){\vector(-1,0){20}}\put(-10,60){\vector(1,0){15}}

\put(10,20){\vector(0,1){35}} \put(55,60){\line(1,0){60}}
\put(115,60){\vector(1,0){30}}\put(15,60){\line(1,0){40}}
\put(135,20){\line(0,1){40}}\put(135, 60){\circle*{1}}

\put(-6,63){\makebox(0,0){$R(s)$}}\put(138,63){\makebox(0,0){$Y(s)$}}
\put(2,63){\makebox(0,0){$+$}}\put(7,52){\makebox(0,0){$+$}}
\end{picture}
\caption{Base element (\ref{elementabL}) as a feedback loop, $s=\rm{i}b$.} \label{elementLK}
\end{center}
\end{figure}
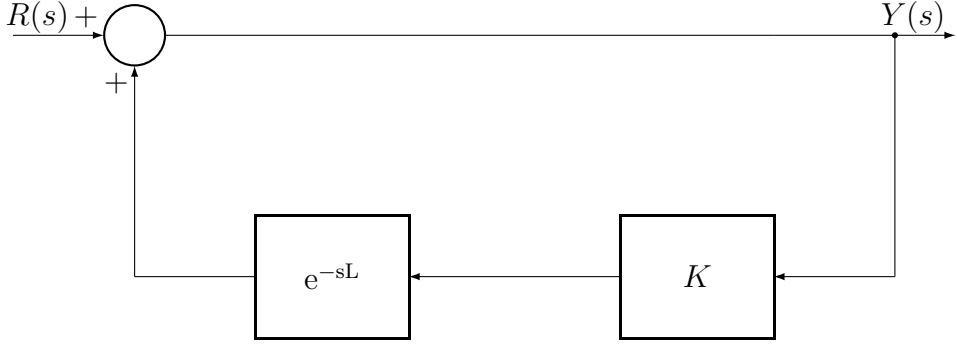

\vspace{0.3cm}\noindent Consider $s=a+\rm{i}b$ for non--negative real $a,b$. Thus, each base element of the zeta function (\ref{elemente}) equals

$$
\dfrac{1}{1-\rm{e}^{-(a+\rm{i}b) \ln p}} 
$$
and by rearranging

\begin{equation}
\dfrac{1}{1-\rm{e}^{-aL}\rm{e}^{-\rm{i}bL}}, \label{elementabL}
\end{equation}
in which $L=\ln p$. Introducing $K=\rm{e}^{-aL}$ as form of a gain, it composes a feedback loop with positive feedback shown in Fig. \ref{elementLK}. Analogously with Fig. \ref{elementL}, it relates to transfer function $Y(s)/R(s)=1/(1-K\rm{e}^{-sL})$ for $s=\rm{i}b$.  

\vspace{0.3cm}\noindent Consider again $L=\ln 2$ for time delay in Fig. \ref{elementLK}. If a reference unit step input occurs, then it immediately is transferred to the output, i.e. $y (t) = 1$ for $t> 0$. In feedback, it is multiplied by gain $K$, delayed by time $L$, and added to the reference input, i.e. $y(t) = 1 + K$ for $t> L$. It again is multiplied by $K$, delayed by time $L$, added to reference input, which products $y(t)=1+K (1 + K) = 1 + K + K ^2$ for $t>2L$, $y(t)=1+K(1+K(1+K))=1+K+K^2+K^3$ for $t>3L$ etc. It forms transient response which is shown in Fig. \ref{zetatK}. Note that a specific case of this response for $K=1$ is in Fig. \ref{zetat} . A complex step signal $y(t)$ is generated by an irregular height of individual steps. In Fig. \ref{zetatK}, point $1+ K^\ast$ on vertical axis relates to $1+K$, $1+2K^\ast$ to $1+K+K^2$, $1+3K^\ast$ to $1+K+K+K^2+K^3$ etc. 

\begin{figure}[htb] 
\begin{center}
\setlength{\unitlength}{1.0mm}
\begin{picture}(100,85)
\thinlines 
\put(10,10){\line(1,0){80}}
\put(10,10){\line(0,1){70}}

\thicklines
\put(10,20){\line(1,0){7}}
\put(17,20){\line(0,1){9}}
\put(17,29){\line(1,0){7}}
\put(24,29){\line(0,1){8}}
\put(24,37){\line(1,0){7}}
\put(31,37){\line(0,1){7}}
\put(31,44){\line(1,0){7}}
\put(38,44){\line(0,1){7}}
\put(38,51){\line(1,0){7}}
\put(45,51){\line(0,1){6}}
\put(45,57){\line(1,0){7}}
\put(52,57){\line(0,1){6}}
\put(52,63){\line(1,0){7}}
\put(59,63){\line(0,1){5}}
\put(59,68){\line(1,0){7}}

\put(17, 29){\circle*{1}}

\thinlines

\put(1,20){\makebox(0,0){$1$}}
\put(1,29){\makebox(0,0){$1+K^\ast$}}
\put(1,37){\makebox(0,0){$1+2K^\ast$}}
\put(1,44){\makebox(0,0){$1+3K^\ast$}}
\put(1,51){\makebox(0,0){$1+4K^\ast$}}
\put(1,57){\makebox(0,0){$1+5K^\ast$}}
\put(1,63){\makebox(0,0){$1+6K^\ast$}}
\put(1,68){\makebox(0,0){$1+7K^\ast$}}

\put(17,7){\makebox(0,0){\small $\ln 2$}}
\put(32,7){\makebox(0,0){\small $3\ln 2$}}
\put(47,7){\makebox(0,0){\small $5\ln 2$}}
\put(62,7){\makebox(0,0){\small $7\ln 2$}}

\put(7,75){\makebox(0,0){$y$}}
\put(80,7){\makebox(0,0){\small $t$}}
\put(10,7){\makebox(0,0){\small $0$}}

\end{picture}
\caption{Transient response associated with transfer function $1/(1-K\rm{e}^{-s \ln 2})$.} \label{zetatK}
\end{center}
\end{figure}

\vspace{-0.3cm}\noindent Since $K\leq 1$ and $K>0$, summation $1+K+K^2+\ldots+K^n$ constitutes  geometric series having partial sums $1+K+K^2+\ldots+K^n=(1-K^{n+1})/(1-K)$ and the infinite sum $1/(1-K)$. A few single time points in Fig. \ref{zetatK} together with related values of transient response is collected in Tab. \ref{zetavalK}.

\begin{table}[htb]
\begin{center}
\begin{tabular}{c|c|c|c|c|c}
$n$ & $1$ & $2$ & $3$ & $4$ & $\ldots$\\ \hline
$y$ & $\dfrac{1-K^2}{1-K}$ & $\dfrac{1-K^3}{1-K}$ & $\dfrac{1-K^4}{1-K}$ & $\dfrac{1-K^5}{1-K}$ & $\ldots$\\
\end{tabular}
\caption{\small Values of transient response representing transfer function $1/(1-K\rm{e}^{-s \ln 2})$.} \label{zetavalK}
\end{center}
\end{table} 

\vspace{0.3cm}\noindent Since a close relation between geometric series and exponential functions, the partial sum of the geometric series can be written as

\begin{equation}
\dfrac{1}{1-K}\left(1-K^{n+1}\right)=\dfrac{1}{1-K}\left(1-\rm{e}^{(n+1)\ln K}\right), \label{geosum}
\end{equation}
in which $n\geq 0$ and $n=0$ relates to $t=0$, $n=1$ to $t=\ln 2$, $n=2$ to $2\ln 2$ etc. 
It expresses transient response of the first order transfer function

\begin{equation}
1+\dfrac{\overline{K}\rm{e}^{-s\overline{L}}}{1+Ts}, \label{gs}
\end{equation}
where $\overline{K}$ is gain, $T$ time constant and $\overline{L}$ time delay. This kind of transfer function is called FOPDT model or three--parameter model \cite{klagor,kla1}. Gain of transfer function (\ref{gs}) is given by the steady state of unit step response $\overline{K}=1/(1-K)-1$ owing to the initial condition $y(0)=1$, then

\begin{equation}
\overline{K}=\dfrac{K}{1-K}.  \label{kp}
\end{equation} 
Identification of the time constant $T$ is based on pairs in Tab. \ref{zetavalK} giving
 
\begin{equation}
T=\dfrac{\ln 2}{1-K}.  \label{tauar}
\end{equation}
It is based on scenario, in which the tangent line of transient response at beginning intersects the point $K$ indicated in Fig. \ref{zetavalK}. If the ratio of $K/(1-K)$ and time constant $T$ determines $\dot{y}(0)$, then the latter is determined by the ratio $K$ and $\ln 2$ as well.\\ 
\\
This approximation suggests that $\overline{L}=0$ permanently for all time delays of FOPDT model (\ref{geosum}). A deeper analysis of transient responses for large prime numbers associated with big delays $L$ and small gains $K$ shows that the shape of the response in Fig. \ref{zetatK} is relatively rare. More often, responses are similar to single steps characterized by delay $L$ and one magnitude $1+K$ since the very small squares of $K$ (\ref{geosum}). This observation was independently confirmed by a use of the three--parameter model calculus \cite{kla1}, which underlines the importance of the time delay approximation in describing behavior of feedback loop in Fig. \ref{elementLK}. Use of this calculus also showed that the time constant (\ref{tauar}) estimates rather give the average residence time of FOPDT model $T_{\rm{ar}}= T+\overline{L}$, see (\ref{gs}) than the time constants themselves. Thus, the estimated time constant results in

\begin{equation}
T=\dfrac{\ln 2}{1-K}\sqrt{K}  \label{tau}
\end{equation}
and the estimated time delay 

\begin{equation}
\overline{L}=\dfrac{\ln 2}{1-K}\left(1-\sqrt{K}\right).  \label{overL}
\end{equation}
These estimates agree with observations. If $K$ is very small, then the estimated time constant is tiny. Conversely, the time delay rises up. Response (\ref{geosum}) looks almost like a single step change in this case.\\
\\ 
Therefore, one can conclude that the Riemann zeta function $\zeta (a + \rm{i} b)$ is described by the endless transfer function called {\it FOPDT approximation} or complex approximation in the form

\begin{equation}
\prod_{p}\left(1+\dfrac{\overline{K}_p \rm{e}^{-s\overline{L_p}}}{1+T_p s}\right), \label{zetag}
\end{equation}
where 

$$K_p=\rm{e}^{-a\ln p}=p^{-a},$$

$$\overline{K}_p=K_p/(1-K_p),$$ 

$$T_p=\dfrac{\ln p}{1-K_p}\sqrt{K_p} $$
and

$$\overline{L_p}=\dfrac{\ln p}{1-K_p}\left(1-\sqrt{K_p}\right) $$
or by the FOPDT frequency function $(\ref{zetag})$

\begin{equation}
\zeta_{\rm{FOPDT}}(a+\rm{i}b)=\prod_{p}\left(1+\dfrac{\overline{K}_p \rm{e}^{-\rm{i}b\overline{L_p}}}{1+\rm{i}bT_p}\right). \label{zetagf}
\end{equation}
Note for completeness, that for Pade approximation of (\ref{elementabL}) we obtain $1/[1-K (1-SL/2)/(1+sL/2)]=(1+sL/2)/[(1-K)+(1+K)sL/2]$, which eventually leads to transfer function based on the single base elements

$$
\dfrac{1}{1-K_p}\ \ \dfrac{1+s \ln p/2}{1+\left(\dfrac{1+K_p}{1-K_p}\ln p/2\right)s}.
$$
This approximation indicates a different initial point for the transient response. Whilst $y (0)=1$ for (\ref{zetag}), $y(0)=1/(1+K_p)$ at Pade approximation. If $a = 0$, then $y(0)=0.5$ in accordance with (\ref{zetapi}). However, the Pade approximation is only applicable to the base elements having small time delays.

\section{FOPDT approximation experiments} 

\vspace{0.3cm}\noindent Some typical real values ($b = 0$) obtained by FOPDT approximation (\ref{zetag}) are collected in Tab. \ref{zetataa}. Experiments were carried out with the first million of prime numbers \cite{cal}. 

\begin{table}[htb]
\begin{center}
\begin{tabular}{r|c|c|r|c|c}
$a$ & $\zeta(a)$ &Tabular value& $a$ & $\zeta(a)$ & Tabular value\\ \hline
$2$ & $1.6449$ & $\pi^2/6$ & $7$ & $1.0083$ &\\
$3$ & $1.2021$ & & $8$ & $1.0041$ & $\pi^8/9450$\\
$4$ & $1.0823$ &$\pi^4/90$ & $9$ & $1.0020$ &\\
$5$ & $1.0369$ & & $10$ & $1.0010$ & $\pi^{10}/93\,555$\\
$6$ & $1.0173$ &$\pi^6/945$ & $12$ & $1.0002$ & $691\pi^{12}/638\,512\,875$\\
\end{tabular}
\caption{\small Some real values of the Riemann zeta function obtained through FOPDT approximation.} \label{zetataa}
\end{center}
\end{table} 

\vspace{0.1cm}\noindent Results for complex values of $s = 2 + \rm{b i}$ obtained by FOPDT approximation (\ref{zetagf}) summarize Tab. \ref{zetataab}. Experiments were carried out with the first million primes as well. For comparison, Tab. \ref{zetataab} refers to tabular values of $\zeta(2+\rm{i}b)$ on the first $4$ decimal places.

\begin{table}[htb]
\begin{center}
\begin{tabular}{r|c|c|r|c|c}
$b$ & $\zeta(2+\rm{i}b)$ &Tabular value& $b$ & $\zeta(2+\rm{i}b)$ & Tabular value\\ \hline
$0.1$ & $1.6351 - 0.0928i$ & $1.6350 - 0.0927i$ & $0.6$ & $1.3807 - 0.4041i$ & $1.3821 - 0.4037i$\\
$0.2$ & $1.6067 - 0.1799i$ & $1.6067 - 0.1798i$ & $0.7$ & $1.3165 - 0.4262i$ & $1.3187 - 0.4261i$\\
$0.3$ & $1.5627 - 0.2567i$ & $1.5628 - 0.2565i$ & $0.8$ & $1.2552 - 0.4375i$ & $1.2582 - 0.4378i$\\
$0.4$ & $1.5075 - 0.3202i$ & $1.5079 - 0.3198i$ & $0.9$ & $1.1980 - 0.4399i$ & $1.2018 - 0.4410i$\\
$0.5$ & $1.4455 - 0.3692i$ & $1.4463 - 0.3688i$ & $1.0$ & $1.1459 - 0.4353i$ & $1.1504 - 0.4375i$\\
\end{tabular}
\caption{\small Some complex values of the zeta function obtained through FOPDT approximation.} \label{zetataab}
\end{center}
\end{table} 

\vspace{0.3cm}\noindent The tables \ref{zetataa} and \ref{zetataab} show that real and complex values of the Riemann zeta function obtained by FOPDT approximation (\ref{zetagf}) are rather accurate. In the case of complex values surprisingly up to two decimal places since achieved by simple FOPDT model. 

\section{Conclusion} 

\vspace{0.3cm}\noindent Based on the similarity of transient characteristics, two approximations of the Riemann zeta function for $a>1$, $\zeta_{\rm {PI}}$ (see (\ref{zetapif})) and $\zeta_{\rm{FOPDT}}$ (see (\ref {zetagf})), were obtained. Approximation of the Riemann zeta function results in a dynamic model consisting of base elements including positive feedbacks and delays, which are infinitely many times serially connected. It provides rather accurate approximations owing to their simplicity.\\  
\\
The paper points to the possibility of using the methods of cybernetics for solving problems in number theory. Two cybernetic--like models were derived close to the Riemann zeta function in terms of single transfer functions. Based on this paper it can be remarkably concluded, that such fundamental concepts of cybernetics as PI controllers or FOPDT models stand in the modeling of the fundamental and complex function as the zeta function undoubtedly is. Can cybernetics be used e.g. in the density analysis of primes with respect to the known role of the Riemann zeta function in this subject?

\end{document}